\documentclass[12pt,epsfig]{article}
\usepackage{graphicx}
\newcommand{\be}{\begin{equation}}
\newcommand{\ee}{\end{equation}}
\newcommand{\gsim}{\, \raisebox{-0.8ex}{$\stackrel{\textstyle >}{\sim}$ }}

\newcommand{\roughly}[1]%
{\mathrel{\raise.4ex\hbox{$#1$\kern-.75em\lower1ex\hbox{$\sim$}}}} 

\newcommand\CL{{\cal L}}

\newcommand\wt{\tilde} 
 
\newcommand\half{\frac{1}{2}} 
\newcommand\beq{\begin{eqnarray}} 
\newcommand\eeq{\end{eqnarray}} 
\newcommand\eqn[1]{\label{eq:#1}} 
 
\newcommand\eq[1]{eq.~(\ref{eq:#1})}

\def\Dsl{\,\raise.15ex \hbox{/}\mkern-12.8mu D} 
\newcommand\Tr{{\rm Tr\,}}

\def\fm3{fm$^{-3}$}

\begin{document}
\centerline{\Large Novel phases and transitions in Color Flavor Locked 
matter}
\bigskip
\centerline{David B. Kaplan\footnote{{\tt
        dbkaplan@phys.washington.edu}} and Sanjay
      Reddy\footnote{{\tt reddy@phys.washington.edu}} } 
\medskip
\centerline{\sl Institute for Nuclear Theory}
\centerline{\sl  University of Washington, 
Box 351550, Seattle, WA 98195-1550, USA }

\begin{abstract} 
We analyze the phases of dense quark matter at nonzero quark masses. We map out
the phase diagrams for charged quark matter as well as charge neutral matter
containing leptons.  We find that in Color Flavor Locked (CFL) quark matter, 
the symmetric phase with equal numbers of $u$, $d$ and $s$ quarks is unlikely 
to occur in nature.  Various less symmetric phases with additional meson
condensates may play a role in neutron star cores, however, in agreement with
the analysis of Bedaque and Sch\"afer. We discuss the implications of these
novel color superconducting quark matter phases and their transitions for
neutron stars.
\end{abstract}

\newpage

\section{The meson condensed phases}

There has recently been renewed interest in the properties of dense quark
matter \cite{thefirst,ALL98}.  Three flavor QCD with massless
quarks at high baryon density is thought to have a particularly symmetric
ground state, the Color Flavor Locked (CFL) phase \cite{Alford:1999mk}. In this
phase the ${ SU(3)_{\rm color}} \times SU(3)_L \times SU(3)_R
\times U(1)_B$ symmetry of QCD is broken down to the global diagonal
$SU(3)$ symmetry due to BCS-like pairing between quarks near the Fermi
surface\footnote{${SU(3)_{\rm color}}$ is the gauged symmetry of QCD; ${
SU(3)_L \times SU(3)_R}$ is the global symmetry which acts on the three flavors
of left- and right-handed quarks; ${ U(1)_B}$ is the baryon number symmetry.}.
As a result, the gluons become massive via the Higgs mechanism. There is the
possibility that color superconductivity is relevant to the structure of
neutron stars. However, in realistic situations, quark masses are nonzero and
there may be nonzero chemical potentials for charge or lepton number, and the
exact flavor symmetry of the QCD Lagrangian is reduced to $U(1)_{em}\times
U(1)_Y\times U(1)_B$, in which case one might expect the CFL condensate to
preserve the $U(1)_{em}\times U(1)_Y$ subgroup of $SU(3)$ not explicitly
broken\footnote{Throughout this paper we will refer to both the $SU(3)$
symmetric phase in the absence of quark masses and chemical potentials, and to
the $U(1)_{em}\times U(1)_Y$ symmetric phase in the presence of masses and
chemical potentials, as the ``symmetric CFL phase''. The identifying
characteristic of the symmetric CFL ground state is that it contains equal
numbers of $u$, $d$ and $s$ \cite{Rajagopal:2001ff}.}.  One can best understand
how the symmetric CFL phase distorts under these applied stresses by studying
the effective theory for the light excitations \cite{Hong:1999dk,
Casalbuoni:1999wu,Son:2000cm,Rho:2000xf,Hong:2000ei,
Manuel:2000wm,Beane:2000ms}, the subject of this Letter.  In particular, we
explore the phase diagram for realistic quark masses in a density regime that
may occur in neutron star interiors, exploiting recent observations by
Sch\"afer and Bedaque
\cite{Bedaque:2001je}. We find that the
symmetric  CFL phase is unstable with respect to meson condensation,
except above baryon densities approximately $10^{21}$ times that
which might be expected in a neutron star core.  Meson condensation results 
in a number
of less symmetric color superconducting phases that may be relevant, with
implications for the thermodynamic and transport properties of neutron star
cores.

The low lying spectrum of the symmetric CFL phase with massless quarks consists
of a nonet of massless Goldstone bosons transforming under the unbroken $SU(3)$
as an octet plus a singlet. A tenth light meson is also expected, a
pseudo-Goldstone boson arising from spontaneously broken axial $U(1)_A$ --- an
approximate symmetry broken by instanton effects, which are expected to be weak
at high density. The massless mesons allow the system to respond continuously
to applied stresses, such as the imposition of nonzero chemical potentials for
$SU(3)$ or $U(1)_B$ currents, since there is no gap in these channels for
massless quarks.

Turning on nonzero quark masses induces a gap in the spectrum. This would seem
to make the symmetric CFL phase rigid against distortion due to applied
chemical potentials.  However, as recently pointed out by Sch\"afer and Bedaque
\cite{Bedaque:2001je}, while quark masses produce gaps in the
spectrum, they also induce new stresses on the system, acting in a manner
analogous to an applied chemical potential and favoring meson condensation.
As discussed by those authors, nonzero quark masses can be expected to lead to
meson condensation and a complicated phase diagram.

We can parameterize the effective theory describing excitations about the
$SU(3)$ symmetric CFL ground state in terms of the two fields $B=e^{i \beta /
f_B}$ and $\Sigma=e^{2i({ \pi}/f_\pi+\eta'/f_A)}$, representing the Goldstone
bosons of broken baryon number $\beta$, and of broken chiral symmetry, the
pseudoscalar octet ${ \pi}$, and the pseudo-Goldstone boson $\eta'$, arising
from broken approximate $U(1)_A$ symmetry. Color degrees of freedom have been
integrated out, and so under the original global $SU(3)_L\times SU(3)_R\times
U(1)_B\times U(1)_A$ symmetry, these fields transform as the representations $
B = (1,1)_{1,0}$ and $\Sigma=(3,\bar 3)_{0,-4} $.

The quark mass matrix $M={\rm diag}(m_u,\, m_d,\, m_s)$ acts like a spurion,
transforming as the representation $( 3,\bar 3)_{0,2}$. Due to the $U(1)_A$
charges, operators in the effective theory must involve even powers of $M$.
Although there are two independent operators one can construct with two powers
of $M$, the leading operator at weak coupling involves only the combination
$\wt M\equiv |M| M^{-1}={\rm diag}(m_d m_s,\, m_s m_u,\, m_u m_d)$, which
transforms as a $( \bar 3, 3)_{0,4}$ and can couple to $\Sigma$ as $\Tr \wt M
\Sigma$.  It is this operator that gives the leading contribution to
the meson masses.  However, one can also consider operators involving
one power of $M$ and one power of $ M^\dagger$.  For example,
the traceless part of $M^\dagger M$, transforms as $(8,1)_{0,0}$,
the same as a left-handed flavor current.  
It has recently been pointed out by Bedaque and Sch\"afer that the combination
$M^\dagger M/(2\mu)$ (where $\mu$ is the quark number chemical potential)
enters the effective theory exactly as would a chemical potential for flavor
symmetries \cite{Bedaque:2001je} \footnote{A similar observation was made 
in Ref. \cite{Manuel:2000wm}}.  As they discuss, this term
destabilizes the ground state for sufficiently large strange quark mass
and leads to kaon condensation.

The leading terms in the effective Lagrangian describing spatially constant
nonet Goldstone boson fields $\pi$ and $\eta'$ in the presence of a chemical
potential $\mu_Q$ for electric charge is given by
\beq
\eqn{leff}
\CL &=& f_{\pi}^2 \left[\frac{1}{4}\Tr D_0 \Sigma D_0 \Sigma^\dagger  
+\half \frac{f_A^2}{f_{\pi}^2} (D_0 \eta')^2 + \frac{a}{2} \Tr
\tilde{M} \left(\Sigma + \Sigma^\dagger\right) + \frac{b}{2}\Tr Q
\Sigma Q \Sigma^\dagger\right]\ \, \nonumber \\ 
& & D_0\Sigma = \partial_0 \Sigma - i\left[
\left(\mu_Q Q + X_L\right)\Sigma -  \Sigma\left(\mu_Q
Q + X_R\right)\right] \,.
\eeq

The decay constants $f_\pi$  and $f_A$ have been computed previously
\cite{Son:2000cm}. $Q$ is
the electric charge matrix ${\rm diag}(2/3,-1/3,-1/3)$ while $X_{L,R}$ are
the Bedaque-Sch\"afer terms: $X_L = -\frac{M M^\dagger}{2\mu}\ ,\quad 
X_R = -\frac{ M^\dagger M}{2\mu}\ \,, $ and $\tilde{M}= |M|M^{-1}$.
   
The leading contributions to meson masses are the $a$ and $b$
operators. The coefficient $a$ has been computed and 
is given by $a=3\frac{\Delta^2}{\pi^2 f_{\pi}^2}$ 
\cite{Son:2000cm} (for a discussion of possible log corrections see 
\cite{Hong:2000ei,Manuel:2000wm,Beane:2000ms}).  The $b$ term accounts 
for electromagnetic corrections to the charged meson masses. The coefficient
$b$ is proportional to $\tilde\alpha$ (the fine structure constant for the
massless photon in the $SU(3)$ symmetric CFL phase) and its value is scheme
dependent\footnote{If one uses a mass independent subtraction scheme in Landau
gauge, the one-loop electromagnetic corrections to the meson masses vanish, and
the $b$ term alone accounts for the entire electromagnetic splittings. We
assume this scheme, and do not include explicit photon fields.}. From naive
dimensional analysis, one can estimate the contribution to $b$ from photon
loops in the effective theory, leading to $ b\sim \frac{\tilde \alpha}{4\pi} 
\Delta^2\ ;$ this formula reflects the quadratic divergence of a photon 
loop in the effective theory, which has a momentum cutoff equal to the
gap $\Delta$.  This estimate is consistent with that of
\cite{Manuel:2001xt}, given that that reference did not attempt to
estimate the factors of $4\pi$ . If the correct value of $b$ differs
from the crude estimate given above, our results will be changed
quantitatively  but not qualitatively.

The meson masses in terms of the parameters $a$, $b$ are
\beq
m^2_{\pi^-} &=& a (m_u + m_d)m_s + b \cr m^2_{K^-} &=& a (m_u + m_s)m_d + b\cr
m^2_{K^0} &=& a (m_d + m_s)m_u \,.
\eqn{masses}
\eeq

The free energy density corresponding to the
Lagrangian in \eq{leff} in the mean field approximation,
relative to the free energy density of the symmetric CFL phase (corresponding to
$\Sigma={\bf 1}$) is:
\beq
\label{omega}
\Omega = \frac{f_{\pi}^2}{4}\left[\Tr [\tilde\mu,\Sigma][ \tilde\mu,\Sigma^\dagger] - 2~a~
 \Tr \wt M (\Sigma  + \Sigma^\dagger-2) - b~\Tr [Q,
  \Sigma][ Q ,\Sigma^\dagger]\right] \ ,\cr 
\eeq
where $\Sigma$ is a constant U(3) matrix characterizing the ground state.  Working in the standard basis where
the quark mass matrix is real, the matrix $\tilde \mu$ is given by
\beq
\tilde \mu = \mu_Q + X\ ,\qquad X=X_L=X_R=-\frac{M^2}{2\mu}\ .
\eeq
The effective chemical potentials vanish for the $\pi^0$, $\eta$ and $\eta'$,
while for the $\pi^+$, ${\rm K}^+$ and ${\rm K}^0$ mesons they are
\beq
\tilde \mu_{\pi^+} = \mu_Q + \frac{m_d^2 - m_u^2}{2\mu}\ ,\quad
\tilde \mu_{K^+} = \mu_Q + \frac{m_s^2 - m_u^2}{2\mu}\ , \quad
\tilde \mu_{K^0} = \frac{m_s^2 - m_d^2}{2\mu}\ .
\eeq

The stationary points of the free energy $\Omega$ with respect to
variations of the meson fields are found as solutions to the matrix
equation 
\beq 
\left[\tilde{\mu}\Sigma^\dagger \tilde{\mu} \Sigma - a\,
M\Sigma - b\, Q \Sigma^\dagger Q \Sigma\right] -h.c.=0 \,.
\eqn{sigmin} 
\eeq 
To find solutions to this equation, we begin by considering $\Sigma$
matrices corresponding to a condensate of a single meson
$\langle\phi\rangle \equiv\sqrt{2}f_\pi\theta $, where $\phi$ could be
a charged pion, a charged kaon, or a neutral kaon field.  Solving
\eq{sigmin} yields stationary points with $\theta\ne 0$ which for
certain ranges of parameters have lower free energy than the symmetric
CFL phase $(\theta=0)$.  We find the solutions \beq
\begin{array}{rlrl}
\Omega_{\pi^\pm} =& -\frac{f_\pi^2}{2} (\tilde \mu_{\pi^\pm}^2
-b)(1-\cos\theta_{\pi^\pm})^2\ ,\quad&\cos\theta_{\pi^\pm}=&\cases{1 &
  $M_{\pi^\pm}^2 \ge \tilde \mu_{\pi^\pm}^2$\cr & \cr
  \frac{M_{\pi^\pm}^2 -b}{\tilde \mu_{\pi^\pm}^2-b}&  $M_{\pi^\pm}^2 \le
  \tilde \mu_{\pi^\pm}^2$}\cr &&\cr
\Omega_{K^\pm} =& -\frac{f_\pi^2}{2} (\tilde \mu_{K^\pm}^2
-b)(1-\cos\theta_{K^\pm})^2\ ,\quad&\cos\theta_{K^\pm}=&\cases{1 &
  $M_{K^\pm}^2 \ge \tilde \mu_{K^\pm}^2$\cr & \cr
  \frac{M_{K^\pm}^2 -b}{\tilde \mu_{K^\pm}^2-b}&  $M_{K^\pm}^2 \le
  \tilde \mu_{K^\pm}^2$}\cr&&\cr
\Omega_{K^0} =& -\frac{f_\pi^2}{2} \tilde \mu_{K^0}^2
(1-\cos\theta_{K^0})^2\ ,\quad&\cos\theta_{K^0}=&\cases{1 &
  $M_{K^0}^2 \ge \tilde \mu_{K^0}^2$\cr & \cr
  \frac{M_{K^0}^2 }{\tilde \mu_{K^0}^2}&  $M_{K^0}^2 \le
  \tilde \mu_{K^0}^2$}
\end{array}
\eqn{scons}
\eeq
Evidently the nontrivial solutions with $\theta\ne 0$ represent phases
with lower free energy than the symmetric CFL phase. The charged pion
and kaon phases can have either electric charge, depending on the sign
of appropriate effective chemical potential $\tilde \mu$. We will
refer to these 3-flavor, color superconducting, meson condensed phases
as ${\rm CFL\pi^\pm}$, ${\rm CFLK^\pm}$ and ${\rm CFLK^0}$
respectively, or simply as ${\pi^\pm}$, ${\rm K^\pm}$ and ${\rm K^0}$,
distinguishing them from the more symmetric CFL phase.

\section{The phase diagram}
\label{sec:2}
\begin{figure}[t]
\begin{center}
\includegraphics[width=.85\textwidth,angle=-90]{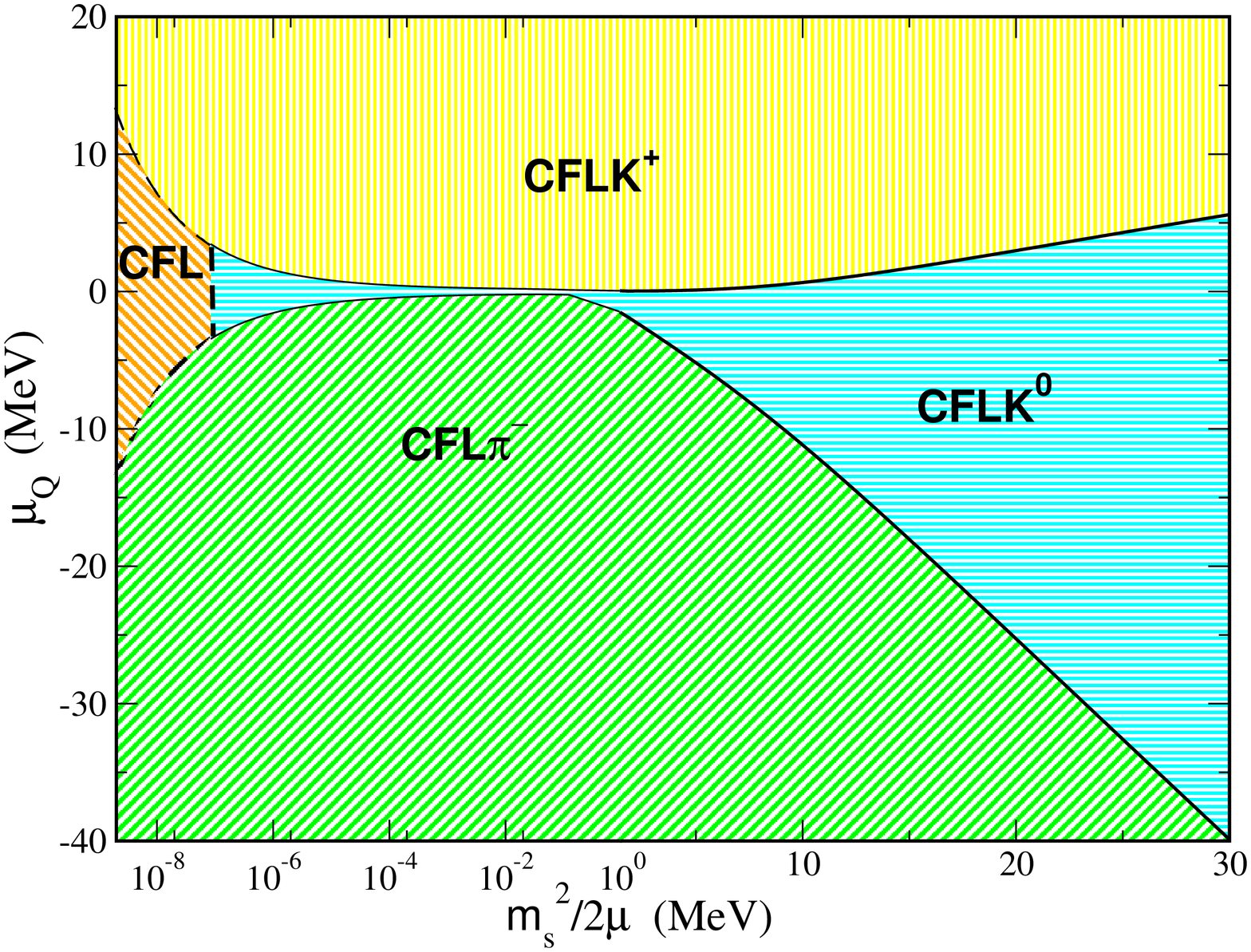}
\end{center}
\caption{Meson condensed phases in the neighborhood of the 
symmetric CFL state are shown in the $(m_s^2/2\mu)-\mu_Q$ plane, where $m_s$ is the
strange quark mass (set to $150$~MeV), $\mu$ is the quark number chemical
potential, and $\mu_Q$ is the chemical potential for positive electric
charge. At five times nuclear density $\mu \sim 400$ MeV and $(m_s^2/2\mu)\sim
25$ MeV. Solid and dashed lines indicate first- and second-order transitions
respectively.}
\label{phase}
\end{figure}

\begin{figure}[t]
\begin{center}
\includegraphics[width=.8\textwidth,angle=-90]{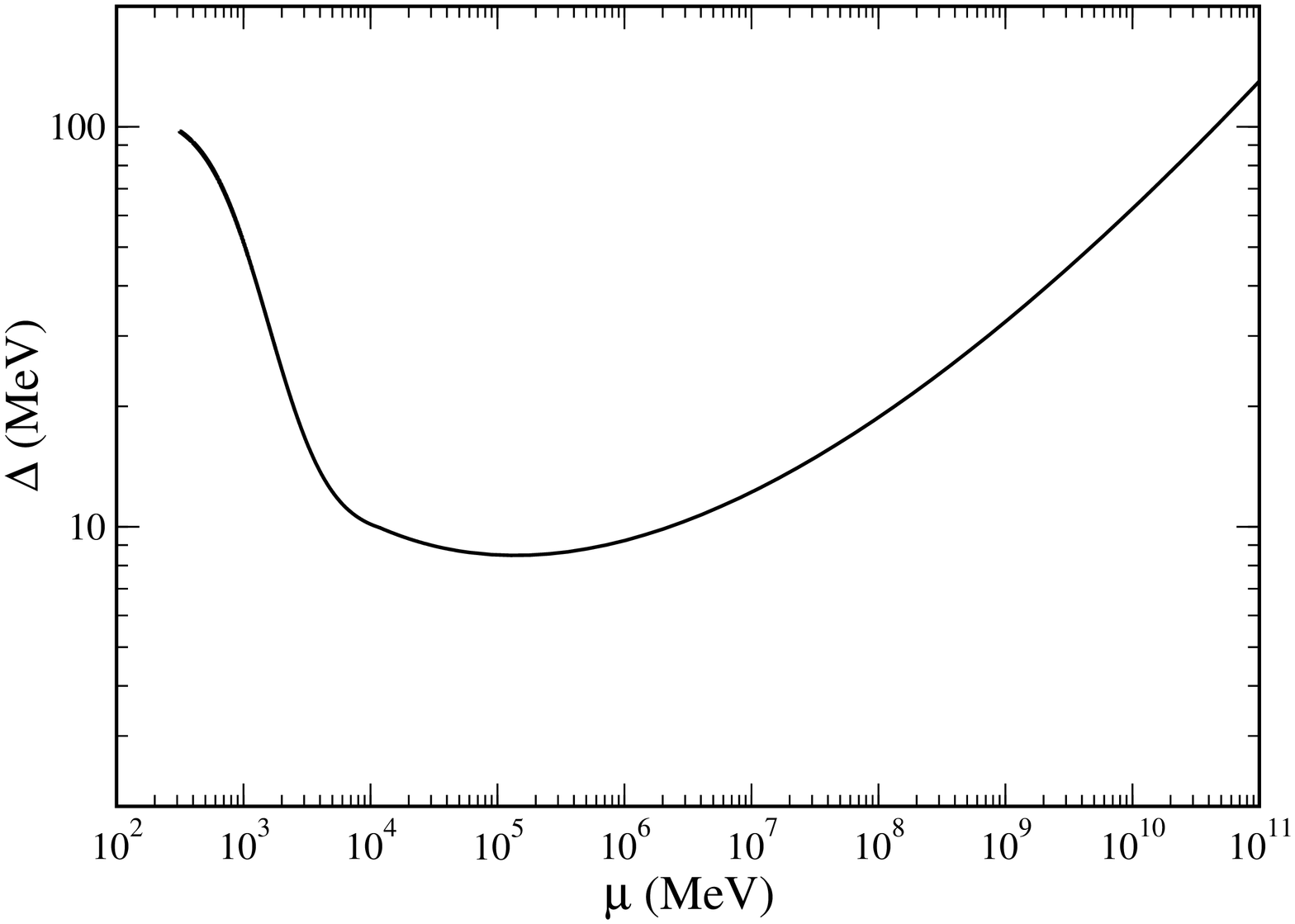}
\end{center}
\caption{A model for superconducting gap $\Delta$ as a function of the
  quark chemical potential $\mu$. The curve agrees with perturbative
  calculations \cite{Son:1999uk} for $\mu>10$~GeV.}
\label{delta}
\end{figure}

In Fig.~1 we plot the phase diagram for color superconducting quark matter, as
a function of $m_s^2/(2\mu)$, where $\mu$ is the quark number chemical
potential, versus the chemical potential for positive electric charge, $\mu_Q$.
In order to construct this diagram we used the quark mass ratios derived from
chiral perturbation theory \cite{Weinberg:1977hb}, $m_u/m_d=1/2$,
$m_d/m_s=1/20$, with the  additional assumption $m_s=150$~MeV.  For the
parameter $b$ in the chiral Lagrangian, we used the rough estimate
$b=\frac{\alpha}{4\pi} \Delta^2$.  Finally, for the relation between the
color superconducting gap $\Delta$ and $\mu$, we used the perturbative
expression from Ref.(\cite{Son:1999uk}) down to $\mu=10$~GeV. At lower
densities, we choose a smooth interpolant that mimics phenomenological models
and predicts $\Delta \sim 100$ MeV for $\mu \sim 500$ MeV
\cite{thefirst}. The superconducting gap obtained in this 
way is shown in Fig.~\ref{delta}. We note that, even in the perturbative
regime, the numerical value of the gap is still not well understood due to
ambiguities associated with scale dependence of the gap equation
\cite{Beane:2001hu}. However, we find that although the precise location of 
the phase boundaries in Fig. \ref{phase} depend on the numerical value of the
gap, the qualitative features of the phase diagram, which will be discussed
below, are fairly independent of $\Delta(\mu)$.

Several features of the phase diagram Fig.~1 are striking.  The first
is that the  
symmetric CFL phase is disfavored until quark densities greater than $\sim
10^{21}$ times what one might find in neutron star cores.  Instead, at
realistic values for $\mu$, the ${\rm K^0}$ phase is favored for small
$\mu_Q$, while the ${\rm K^+}$ and ${\rm \pi^-}$ phases are
  favored for $\mu_Q$ above 
or below critical values. The ${\rm \bar K^0}$ and ${\rm K^-}$
    condensed phases are 
disfavored as they introduce strange quarks in a system trying to reduce their
number\footnote{This is in contrast to kaon condensation in hadronic matter
\cite{Kaplan:1986yq, Nelson:1987dg}, which suffers from a paucity of
strange quarks, while the symmetric CFL phase has a surplus.}, while for
positive $\mu_Q$  ${\rm K^+}$ is favored over  
${\rm \pi^+}$ both because the ${\rm K}^+$ is lighter, and because a ${\rm K}^+$
condensate  reduces the
strange quark number while a pion condensate does not.

As one approaches from either side the boundaries between the ${\rm
  K^0}$ phase and the 
${\rm K^+}$
or ${\rm \pi^-}$ phases, one finds that the meson condensate angles
$\theta$ given in \eq{scons} do not vanish.  This indicates that the
phase transitions are first order.  One can verify this by solving the
stationarity equation \eq{sigmin} with a two condensate ansatz for the
$\Sigma$ matrix.  As an example we analyze the  ${\rm K^+/K^0}$
transition. With the ansatz
\beq
\Sigma = {\rm exp}~\left[i\theta\left(\matrix{0 & 0 & \sin\phi\cr 0 &
      0 & \cos\phi\cr 
  \sin\phi & \cos\phi & 0\cr}\right)\right]
\eeq
one finds a  saddle point
solution to \eq{sigmin} interpolating between the ${\rm K^0}$ and
${\rm K^+}$ phases, given by
\beq
\cos\theta &=& \frac{M_{K^0}^2}{\tilde\mu_{K^0}^2} +
\frac{\tilde\mu_{K^+}(\tilde\mu_{K^+}-\tilde\mu_{K^0})}{b\tilde\mu_{K^0}}
\left[\left(\frac{M_{K^+}^2-\tilde 
      \mu^2_{K^+}}{\tilde\mu_{K^+}}\right) - \left(\frac{M_{K^0}^2-\tilde
      \mu^2_{K^0}}{\tilde\mu_{K^0}}\right)\right]\ ,\cr &&\cr
\sin^2\phi &=& \frac{\left(\frac{M_{K^0}^2}{\tilde
      \mu_{K^0}^2}-\cos\theta\right)}{\left(1-\frac{\tilde\mu_{K^+}}
    {\tilde\mu_{K^0}}\right)(1-\cos\theta)}\ .     
\eqn{spangles}
\eeq
This saddle point solution exists only for ranges of parameters for which the
angles in \eq{spangles} are real. The free energy density of this
 solution relative to that 
of the ${\rm K^0}$ phase  is
\beq
\delta\Omega\equiv \left(\Omega_{K^0/K^+}-\Omega_{K^0}\right) =
\frac{f_\pi^2}{4}~\frac{(m_{K^0}^2-\mu_{k^0}^2\cos{\theta})^2}
{\mu_{K^0}^2~(\mu_{K^0}-\mu_{K^+})^2}~b\,.
\label{barrier}
\eeq
It is the quantity $\delta\Omega$ which will control the tunneling
rate between the ${\rm K^+}$ and ${\rm K^0}$ phases during a
first order phase transition.
 
\section{Neutron stars}
\label{sec:3}

If quark matter were to exist in neutron stars, the typical quark chemical
potential is $\mu \sim 400$ MeV. As discussed earlier, perturbative calculations
are inapplicable here and model calculations based on phenomenological
four-fermion interactions suggest that three flavor massless quark matter may be
characterized by a gap as large as $\Delta \sim 100$ MeV. As will become clear
from the following discussion, this is larger than the typical energy
associated with external stresses that quark matter is subject to in the
neutron star interior. Therefore we can calculate the response of quark matter
to these stresses within the frame work of the effective theory defined by
eq.~(1).

Neutron stars are born in the aftermath of a core collapse supernova
explosion. The inner core of the newly born neutron star is
characterized by a temperature ${\rm T}\sim 30$ MeV and a lepton
fraction (lepton number/baryon number) ${\rm Y_L} \sim 0.3$ implying $\mu_e
\equiv \mu_{\nu_e}-\mu_Q \sim \mu_{\nu_e} \sim 200$~MeV. The high
temperature and finite lepton chemical potentials are a consequence of
neutrino trapping during gravitational collapse
\cite{Burrows:1986me}. Subsequently, the interior cools and looses lepton
number via neutrino diffusion. This occurs on a time scale of few tens of
seconds. At later times, ${\rm T} \ll 1$~MeV, $\mu_{\nu_e}=0$ and the electron
chemical potential is determined by the condition of electric charge
neutrality. In noninteracting quark matter, $\mu_e
\sim m_s^2/2\mu \sim 25$~MeV. However, in the superconducting phase with
$\Delta \gsim m_s^2/4\mu$ our phase diagram Fig.~1 indicates that the
ground state is the ${ \rm K^0}$ phase. The ${\rm K^0}$ phase is
electrically neutral without electrons. This result has far reaching
consequences because, being devoid of electrons, the quark phase has
no low lying charged excitations up to the mass of the lightest
charged Goldstone boson.  In the symmetric CFL phase, neutrality is
guaranteed by the enforced equality of the u, d, and s quarks
\cite{Rajagopal:2001ff}. For the ${ \rm K^0}$ phase, charge neutrality
is a fluke and is due to the fact that the ${\rm K^+}$ is heavier than
the ${\rm K}^0$.

The core of a neutron star during the early stages of its evolution
will be charge neutral, and will have nonzero lepton number. Assuming
local electric charge neutrality, the possible ground states among the
phases we have discussed are ${\rm K^0}$ condensation with all the
leptons being neutrinos; ${\rm K^+}$ condensation with charge
neutralizing electrons; or a mixed ${\rm K^+/K^0}$ condensate with
electrons, corresponding to $0<\phi<\pi/2$ in eq. (8).  Our analysis
at fixed $\mu_Q$ found that the latter phase was not a minimum of the
free energy, but a saddle point.  Remarkably enough, when leptons are
included and local charge neutrality is imposed, the saddle point
solution of eq. (11) becomes the true minimum of the free energy over
the range that it exists.  Further, we find that it smoothly connects
the ${ \rm K^0}$ phase with ${ \rm K^+}$ phase resulting in second
order transition transitions. Note that \eq{sigmin} remains valid and
characterizes the stationary points of the free energy even when
charge neutrality is imposed at fixed lepton chemical potential.

\begin{figure}[t]
\begin{center}
\includegraphics[width=.85\textwidth,angle=-90]{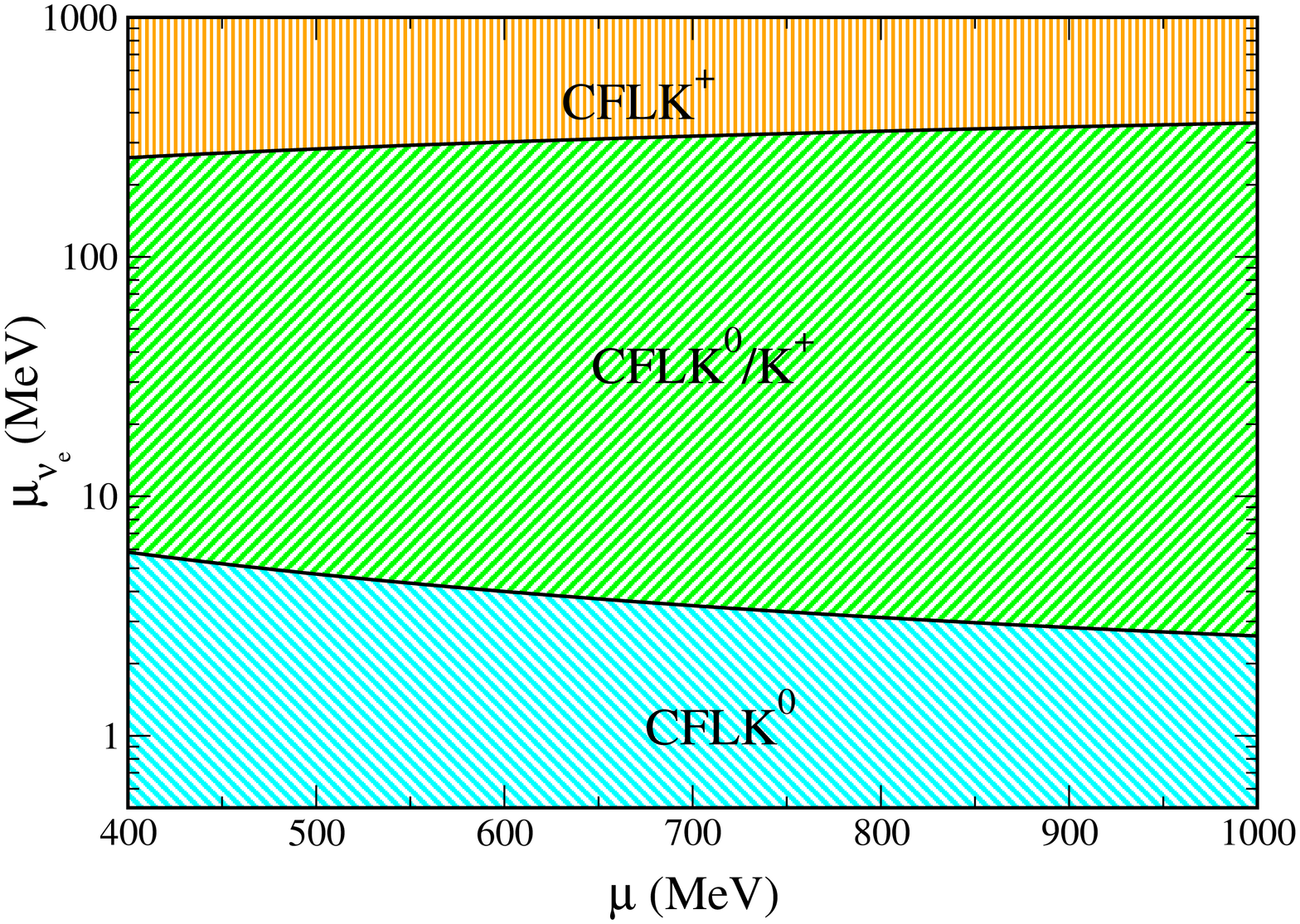}
\end{center}
\caption{The phase diagram for charge neutral matter as a function of
the quark number and neutrino number chemical potentials $\mu$ and
$\mu_{\nu_e}$. The phase transitions shown are both second order.}
\label{neutral}
\end{figure}

A phase diagram depicting each of the three phases discussed above are
shown in Fig.~\ref{neutral}. Electric charge neutrality is enforced in
the phase containing the ${\rm K^+}$ mesons by requiring that the
electron number density $n_e= -\partial \Omega_{{\rm meson}}/\partial
\mu_Q$. This uniquely determines the electron chemical potential and
hence their contribution to the total free energy.  In
Fig.~\ref{neutral} we restrict the quark chemical potential to the
range of relevance for neutron stars: 400-1000 MeV; and make the
simplifying assumption that the gap $\Delta=100$ MeV over this range.

From Fig.~\ref{neutral} we can infer that the ground state will
contain electrons for $\mu_{\nu_e} \gsim 6$ MeV \footnote{ This
depends on the value of b, for example, for $b=e^2\Delta^2$ (16$\pi^2$
times larger than the value we use) electrons appear for $\mu_{\nu_e}
\gsim 20$ MeV} . The superconducting phase admits electrons with ease
since the ${\rm K}^+$ mesons are not much heavier than the neutral
${\rm K}^0$ mesons.  We find that the electric charge chemical
potential $\mu_Q$ is only a few MeV even when $\mu_{\nu_e} \sim 200$
MeV. Therefore the induced stress on the quark phase, characterized by
$\tilde{\mu}_{\rm K^+}=\mu_Q+X$, is always small compared to
$\Delta$. Consequently, unless excluded by high temperatures, quark
matter in neutron star cores will be color superconducting even when
leptons are trapped. As lepton number is lost via neutrino diffusion,
we can expect a second order transition from the ${\rm K^+}$ to the
two condensate phase. Upon further deleptonization, when $\mu_{\nu_e}
\sim 5-10 $ MeV another second order transition from the ${\rm
K^0/K^+}$ phase (containing electrons) to the ${\rm K^0}$ phase
(without electrons) is to be expected.

Although it is premature to speculate precisely how this will
influence various observable aspects of the early evolution of neutron
stars we comment on a few important consequences of our
findings. Firstly, since the only hadronic excitations in the ${\rm
CFL\phi}$ phases with energies below the gap are the pseudo Goldstone
bosons with masses $\sim 10-20$ MeV the leptons are likely to dominate
the transport processes and the matter specific heat. During
deleptonization we can expect dramatic changes to these fore mentioned
properties. This is likely to affect various aspects of the early
evolution of the neutron star including neutrino transport.

If magnetic fields play a role during this early phase, we can expect field
topology to change greatly during the early deleptonization phase since the
initial ${\rm K^+}$ and ${\rm K^0/K^+}$ phases are electrical superconductors
they will expel the magnetic field in the core due to Meissner effect; while at
later times when the system is characterized by a neutral ${\rm K^0}$ phase
and the U(1)$_{\rm em}$ is restored, the Meissner affect disappears and the
core now admits the magnetic field. 

Finally, since the neutrino free ground state contains no electrons, we note
that the total lepton number emitted in neutrinos will be $20\%-30\%$ larger
than in conventional scenarios. This is particularly, exciting because current
neutrino detectors such as SNO, which can distinguish between different
flavors, in conjunction with larger detectors, such as Super Kamiokande which
primarily detect the anti-electron neutrinos, is in principle sensitive to the
lepton number emitted from a galactic supernova.

It must be stressed that there are significant uncertainties in our
analysis.  Most importantly, the density regime one expects in neutron
star cores is not high enough for the application of perturbative QCD.
At these densities, the question of whether a quark phase exists at
all, whether color superconductivity occurs should such a quark phase
exist, and what the size of the resulting gap and chiral Lagrangian
parameters would be, are all presently unanswerable.  In our analysis
we have extrapolated from the perturbative regime and respected the
symmetries of QCD, with the exception that we have assumed $U(1)_A$
violating instanton effects to be small.  If the latter assumptions
proves to be false at the relatively low densities found in neutron
stars, a $\Tr M\Sigma$ will be induced in the chiral Lagrangian 
\cite{Alford:1999mk,Manuel:2000wm},
altering our analysis. We also note that we have ignored finite
temperature effects in our analysis of the phase diagram. Thus we
require that the matter temperature be small compared to the
superconducting critical temperature ${\rm T_c}\sim \Delta$ . In addition,
we have neglected finite temperature corrections to the effective
theory and the role of thermally excited mesons, which we expect to be
small in the protoneutron star context.  However, the contribution of
thermal mesons to the free energy may not be negligible. Nevertheless,
we feel confident that should color superconductivity exist in neutron
stars, the phase diagram we have computed will be qualitatively
correct and warrants further investigation.


\vskip0.75in
\centerline{\bf Acknowledgements}
\bigskip

We thank S. Beane, G. F. Bertsch, K. Rajagopal, M. Savage, and T. Sch\"afer for
useful comments. This work was supported by DOE grant DE-FG03-00-ER-41132.
\vfill
\eject

\end{document}